\documentclass[12pt,twoside]{article}   %For printing on two sides of page
\usepackage[super,sort,comma]{natbib}
\usepackage{fancyhdr}		%Gives headers and footers defined below
\usepackage{enumitem}

\usepackage[section]{placeins}   %
\usepackage{graphicx}
\graphicspath{{figures/}}
\usepackage{booktabs}
\usepackage{tabularx}
\usepackage{threeparttable}
\usepackage{float}

\usepackage{booktabs}
\usepackage{tabularx}
\usepackage{threeparttable}
\usepackage{array}
\usepackage{makecell}
\usepackage{caption}
\usepackage{placeins}

\usepackage{pgfplots}
\pgfplotsset{compat=1.18}

\usepackage{tikz}
\usetikzlibrary{arrows.meta, positioning}

\makeatletter \renewcommand\@biblabel[1]{$^{#1}$} \makeatother
\usepackage{hyperref}
\hypersetup{ colorlinks,
    citecolor=blue,
    filecolor=blue,
    linkcolor=blue,
    urlcolor=blue
}

\usepackage{xcolor}
\definecolor{gray}{rgb}{0.6,0.6,0.6}
\definecolor{red}{rgb}{0.85,0,0}
\definecolor{green}{rgb}{0,0.85,0}
\definecolor{blue}{rgb}{0,0,0.85}
\definecolor{beige}{rgb}{0.92,0.87,0.78}
\usepackage[all]{hypcap}    %causes link to figures to go to figure, not caption
\usepackage{amsmath}
\usepackage{amsfonts}
\usepackage{amssymb}
\usepackage{algorithmic}
\usepackage{algorithm2e}
\usepackage{verbatim}

\newcommand{\revision}{\color{black}}

\begin{document}

\begin{center}
{\Large\bfseries  Evaluating a Large Language Model for Radiotherapy Patient Scheduling under Realistic Clinical and Operational Constraints: a Feasibility Study\par}
\vspace{6pt}

% Eric Zhang$^{1,2}$, Wen Li$^{1}$, Youfang Lai$^{1}$, Annette Souranis$^{1}$, Georgia Paparoidamis$^{1}$,\\
% Michael Roumeliotis$^{1}$, Xun Jia$^{1}$

\vspace{6pt}

Eric Zhang$^{1,2}$, Youfang Lai$^{1}$, Wen Li$^{1}$, Annette Souranis$^{1}$, Georgia Paparoidamis$^{1}$,\\
Michael Roumeliotis$^{1}$, Xun Jia$^{1}$

\vspace{6pt}

{\small
$^{1}$Department of Radiation Oncology and Molecular Radiation Sciences, Johns Hopkins University\\
$^{2}$Westlake High School, Westlake Village, California

}
\end{center}

\vspace{10pt}

\date{}

\begin{abstract}  % <= 300 words
\noindent\textbf{Background:}  {\revision Radiotherapy (RT) scheduling requires coordination of clinical constraints and operational objectives. Large language models (LLMs) may provide a natural-language interface for generating candidate schedules.}
  
\noindent\textbf{Purpose:}  {\revision To evaluate the feasibility of using an existing LLM to generate RT patient schedules under clinical and operational constraints through simulation and retrospective clinical comparison.}

\noindent\textbf{Methods:}  {\revision A one-year, three-linear-accelerator (LINAC) simulation included 1,400 new patients across 12 treatment categories modeled after clinical practice.} Structured prompts supplied ChatGPT with patient data, the current schedule state, clinical rules, and scenario-specific objectives. Performance was evaluated across scenarios addressing weekly treatment-time consistency, LINAC continuity, gap-constrained temporal relaxation, and infeasible request handling. {\revision LLM-generated schedules were benchmarked against heuristic and CP-SAT schedules under two objective settings.} {\revision A six-week retrospective comparison evaluated 2,043 appointments against a clinical schedule.}

\noindent\textbf{Results:}  The LLM-generated schedules satisfied the predefined feasibility rules in all evaluated scenarios. {\revision Across scenarios, 98.6--98.8\% of evaluated fractions remained within the preferred treatment-time window.} Adding a LINAC-continuity objective reduced the LINAC switch rate from 52.4 $\pm$ 2.6\% to 9.4 $\pm$ 0.6\%, and gap-constrained temporal relaxation reduced daily gap to 47.1 $\pm$ 8.5 minutes and Friday gap to 89.2 $\pm$ 18.4 minutes while maintaining low overtime. In infeasible request tests, the framework identified predefined constraint violations and returned interpretable corrective suggestions. The optimization benchmarks also generated feasible schedules, but their results showed strong objective-weight trade-offs. {\revision In the retrospective comparison, all computational approaches scheduled the workload without overtime; the LLM strategies reduced mean gap by 50.8--77.1\% and increased time consistency from 86.7\% to 99.5--99.9\%.}

\noindent\textbf{Conclusion:}  {\revision Structured LLM-assisted scheduling was feasible in the simulation and retrospective clinical comparison, producing candidate schedules under predefined constraints.} The findings support further investigation of LLMs as flexible, human-in-the-loop decision-support tools for translating clinical priorities into feasible candidate schedules.

\end{abstract}


%\maketitle

\pagenumbering{roman}
\setcounter{page}{1}
\pagestyle{plain}

\newpage     %may or may not be needed

%The table of contents is for drafting and refereeing purposes only. Note
%that all links to references, tables and figures can be clicked on and
%returned to calling point using cmd[ on a Mac using Preview or some
%equivalent on PCs (see View - go to on whatever reader).
\tableofcontents

\newpage

\setlength{\baselineskip}{0.7cm}      %double spacing		
\pagenumbering{arabic}
\setcounter{page}{1}
\pagestyle{fancy}

\section{Introduction}

Radiation therapy (RT) patient scheduling represents a complex resource allocation problem. Unlike general outpatient appointment scheduling, RT scheduling must coordinate multi-fraction treatment courses delivered over several weeks while satisfying numerous clinical and operational requirements \citep{kapamara2006review}, including designated time slots for extended first-fraction treatments, minimum spacing requirements for multi-fraction regimens, treatment machine eligibility restrictions, and machine and staffing availability. These requirements directly affect RT resource utilization. Advances in RT practice further emphasize the need for efficient resource planning. For example, the increasing adoption of hypofractionated treatment regimens can substantially alter treatment capacity requirements\citep{roumeliotis2024impact}. Adaptive RT introduces additional scheduling complexity because treatment replanning, physician review, quality assurance must be coordinated within clinically relevant time windows\citep{sonke2019adaptive,li2013automatic}. In current clinical practice, scheduling remains largely manual and is performed by experienced administrators and therapists. This process is labor-intensive, cognitively demanding, and difficult to scale as patient volumes increase or operational conditions change. As a result, manual scheduling may limit efficient use of treatment resources when competing clinical and operational constraints must be balanced simultaneously.

The complexity of the RT scheduling problem has motivated extensive research within operations research frameworks. A comprehensive review highlights the broad application of operations research methods for logistics optimization and resource planning in RT across strategic, tactical, and operational levels \cite{Vieira2016ORReview}. Early optimization-based approaches formulated RT scheduling as a mathematical programming problem, notably through priority-based non-block scheduling models designed to improve treatment allocation while accounting for patient urgency\cite{conforti2010nonblock}. More recent studies have employed exact optimization models, such as integer programming, constraint programming, and column generation \cite{Frimodig2023Comparison,frimodig2026column}. The successful translation of these theoretical operations research models into clinical practice has been demonstrated by implementing frameworks that optimize scheduling while strictly respecting clinical time window preferences\citep{vieira2020timewindow,Vieira2021Implementation}. Meanwhile, to address the inherent unpredictability of patient inflows, researchers have developed stochastic and simulation-based dynamic optimization strategies, showcasing the efficacy of adaptive decision-making under uncertain demand\cite{Legrain2015OnlineStochastic,gocgun2018simulation, Vieira2019DES}. More recently, data-driven and machine learning approaches have gained traction in this domain. Predictive models were employed to estimate RT treatment durations, enabling more efficient workday divisions, better treatment scheduling management and device utilization \citep{Bentayeb2019ServiceTime,xie2023MLTimePrediction}. An online dynamic patient scheduling framework was introduced to learn the relationships between patient features and actual wait times, dynamically adapting scheduling decisions to outperform traditional flat reservation policies \citep{Pham2023PredictionScheduling,khalfi2026ai}. Despite these advances, clinical translation can remain challenging because scheduling objectives, soft constraints, and exception-handling rules must be explicitly encoded and maintained. When local policies change or unexpected operational events occur, optimization models may require reformulation, parameter adjustment, or additional implementation effort.

Recent advances in artificial intelligence (AI) have increasingly influenced medical physics and RT \citep{shen2020introduction,shan2020synergizing, nguyen2022advances}. Within this landscape, large language models (LLMs) have emerged as versatile tools capable of interpreting complex natural language instructions, synthesizing structured data, and executing multi-step reasoning. LLM-based methods have been investigated for tasks such as treatment target delineation, patient-facing question answering, treatment planning, and structure-name standardization \citep{rajendran2025large,yalamanchili2024quality,holmes2024benchmarking,jafar2026human, hao2025personalizing}. Beyond these RT applications, LLMs show promise in automating broader healthcare workflows, including administrative coordination, documentation, and digital task automation \citep{tripathi2024efficient,zhu2025medicalos}.  These capabilities are uniquely suited to the RT scheduling problem, a domain where patient data, institutional policies, and clinical operational preferences must be integrated to yield feasible decisions. In contrast to conventional optimization-based approaches that require formal mathematical encoding, LLMs may provide a complementary interface for translating heterogeneous natural-language policies, structured patient data, and operational preferences into candidate schedules. This inherent flexibility is particularly advantageous for RT scheduling, where operational logic is frequently procedural rather than strictly algebraic. Complex clinical scheduling therefore provides a compelling test case for LLMs. Evaluating an LLM in this setting can help determine whether prompt-guided reasoning can complement optimization-based approaches by translating complex clinical and operational context into feasible schedules.

 In this study, we evaluated the feasibility of using an existing LLM, ChatGPT (OpenAI Inc, San Francisco, CA), to generate candidate RT patient schedules under predefined clinical and operational constraints. To perform this evaluation, we developed a structured prompting framework that provided patient data, existing schedule states, RT scheduling rules, and operational context to the LLM to generate feasible schedules meeting all requirements. {\revision We also compared the LLM-generated schedules with optimization-based scheduling results in a simulation study to provide an initial benchmark for interpreting LLM's performance. We additionally conducted a six-week retrospective comparison of the LLM-based and optimization-based scheduling results with a clinical schedule to assess scheduling performance at routine clinical workload. } To our knowledge, this study represents one of the first systematic evaluations of structured LLM-assisted schedule generation for RT patient scheduling under realistic conditions.

\section{Methods}
\label{sec:method}

\subsection{Clinical Environment, Patient Categories, and Scheduling Constraints}
\label{sec:scheduleconstraints}

The simulated scheduling environment was modeled after the workflow in our RT department with three medical linear accelerators (LINACs).  To define the clinical scheduling workflow, operational constraints, and representative treatment categories, we consulted with our radiation therapist team to identify routine clinical scheduling features including treatment categories, fractionation patterns, LINAC eligibility, first-fraction timing rules, anesthesia and BID constraints, machine operating hours, and variable daily workload etc.

Treatments are delivered within a standard clinical operating window from 7:00 AM to 5:00 PM, corresponding to 600 available minutes per LINAC. Overtime is allowed if necessary to complete treatments. To capture the diversity of clinical treatment workflows, patients were categorized into 12 groups based on treatment modality and scheduling requirements. These categories include stereotactic body radiation therapy (SBRT), breast treatments, total body irradiation (TBI), anesthesia cases, twice-daily (BID) treatments, multi-isocenter treatments, and other fractionated courses. The full set of patient categories, along with machine eligibility, treatment time per fraction, number of fractions, historical patient mix proportions, and scheduling constraints, is summarized in Table~\ref{tab:patient_categories}. Each category is associated with specific operational constraints. For example, BID treatments require a minimum interval of six hours between fractions delivered on the same day, while anesthesia cases must be scheduled at restricted morning time slots due to anesthesia workflow considerations. To maintain physician availability and supervised workflows, new treatment starts are initiated at predefined, machine-specific time slots. These starts occur at 9:00 AM, 11:00 AM, and 2:00 PM for LINAC 1, 9:30 AM, 11:00 AM, and 2:00 PM for LINAC 2; for LINAC 3, the designated slots are 10:00 AM, 1:00 PM, and 2:00 PM. New treatments are not initiated on Fridays to prevent care interruptions over the weekend. Furthermore, for multi-fraction treatment courses, the initial fraction for any new course includes an additional 15 minutes to accommodate comprehensive setup and verification procedures.

These rules delineate the feasibility space of the scheduling problem, serving as the essential constraints that the scheduling framework must satisfy. While these parameters are not intended to reproduce every daily variation, they provide a clinically motivated approximation of routine operations. Consequently, they establish a realistic and robust foundation for the scheduling simulations and performance evaluations presented throughout this paper.  Note that the simulated environment was designed to represent regular scheduling behavior under normal operating conditions, rather than last-minute disruptions such as cancellations, machine downtime, emergency add-ons, or adaptive replanning.

\begin{table*}[t]
\centering
\caption{Patient categories based on treatment scheduling requirements. Abbreviations: Fx = fraction; Tx = treatment; BID = twice daily; TBI = total body irradiation.}
\label{tab:patient_categories}
\small
\setlength{\tabcolsep}{3pt}
\renewcommand{\arraystretch}{1.15}

\begin{tabularx}{6.5in}{
>{\centering\arraybackslash}p{0.45in}
>{\raggedright\arraybackslash}p{1in}
>{\centering\arraybackslash}p{0.55in}
>{\centering\arraybackslash}p{0.55in}
>{\centering\arraybackslash}p{0.55in}
>{\centering\arraybackslash}p{0.65in}
>{\centering\arraybackslash}p{0.75in}
>{\raggedright\arraybackslash}p{1.10in}
}
\hline
\makecell{Cate- \\gory} 
& Name 
& LINAC 
& \makecell{Tx \\duration\\(min)} 
& \makecell{No.\\of Fx} 
& \makecell{Patient\\mix (\%)} 
& \makecell{Days of\\1st Fx} 
& \makecell{Constraints on\\all Fx} \\
\hline
1  & SBRT1              & 3     & 30 & 5  & 10\% & No Fri.   & --- \\
2  & SBRT2              & 3     & 40 & 5  & 5\%  & No Fri.   & --- \\
3  & Other short-course & 3     & 15 & 5  & 5\%  & No Fri.   & --- \\
4  & Breast1            & 1,2   & 15 & 5  & 2\%  & No Fri.   & --- \\
5  & Breast2            & 1,2   & 15 & 16 & 8\%  & Mon. only & --- \\
6  & TBI1               & 1     & 30 & 1  & 1\%  & No Fri.   & --- \\
7  & TBI2               & 1     & 60 & 1  & 7\%  & No Fri.   & --- \\
8  & Anesthesia1        & 1,2,3 & 60 & 1  & 1\%  & No Fri.   & 9am Thu; 8am other days \\
9  & Anesthesia2        & 1,2,3 & 60 & 1  & 1\%  & No Fri.   & Before 12pm \\
10 & BID                & 1,2,3 & 15 & 52 & 5\%  & No Fri.   & $\geq$ 6h apart \\
11 & Multi-iso          & 1,2,3 & 25 & 10 & 5\%  & No Fri.   & --- \\
12 & Other conventional & 1,2,3 & 15 & 25 & 50\% & No Fri.   & --- \\
\hline
\end{tabularx}
\end{table*}

\subsection{Patient Data Generation and Simulation Setup}
\label{sec:datagen}

A synthetic patient dataset was generated to evaluate the scheduling framework under controlled yet clinically representative conditions. The simulation spans a total of 260 treatment days, corresponding to one full clinical year.  The treatment calendar was represented in terms of available operating days rather than every calendar day; planned non-treatment days, such as weekends or predefined clinic closures, were skipped when constructing each patient's treatment course, and subsequent fractions were scheduled on the next available operating day.

To eliminate artificial initialization effects associated with an empty system, the simulation was initialized with the treatment calendar populated by a cohort of ongoing (“existing”) patients on Day 1 sampled from Table \ref{tab:patient_categories}.  Each existing patient was assigned a remaining number of fractions consistent with the corresponding treatment category. A total of 1,400 new patients were then generated over the 260-day horizon. Daily new-patient counts were sampled from a uniform distribution between 5 and 9 on eligible start days. Patient categories were sampled according to the proportions in Table \ref{tab:patient_categories}. Category-specific attributes, including the number of fractions and treatment time per fraction, were assigned accordingly. {\revision The simulated workload was deliberately set higher than routine clinical workload to test scheduling feasibility and competing operational objectives under more constrained capacity.}

The simulation was then executed sequentially on a day-by-day basis to schedule these patients according to their fractionation schedules. On each scheduling day, newly arriving patients were added to the treatment calendar, while ongoing patients continued according to their prescribed fractionation schedules. Patients who completed the treatment course were removed from the calendar.

\begin{figure}[htbp]
\centering
\includegraphics[width = 6 in]{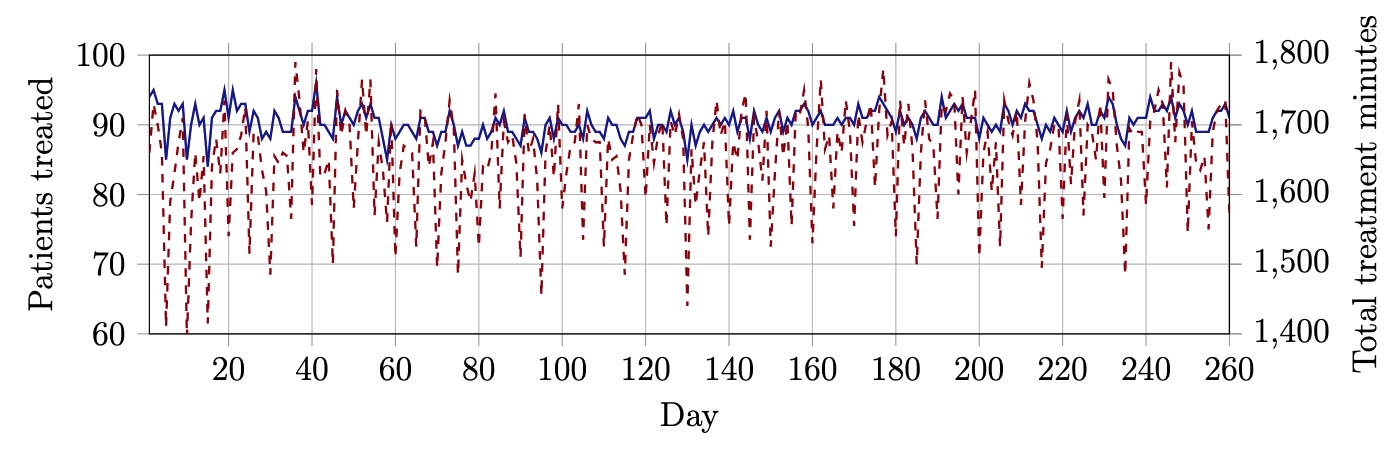}

\caption{Daily number of patients (solid line) and total scheduled treatment time across all LINACs (dashed line) over the 260-day simulation period.}
\label{fig:combined}
\end{figure}

The simulated dataset exhibited stable daily patient volume and machine utilization over the full 260-day period. As shown in Fig.~\ref{fig:combined}, the number of patients treated per day ranges from 84 to 96, with an average of 90.5 patients per day. The corresponding total treatment time varies between 1,395 and 1,790 minutes per day, with a mean of approximately 1,655 minutes across all LINACs.

% Weekly patterns are further illustrated in Fig.~\ref{fig:combined}(b) and  Fig.~\ref{fig:combined}(c), where patient volume and treatment workload are relatively uniform from Monday through Thursday, with a modest reduction on Fridays due to scheduling constraints on new patient starts.

% This dataset provides a realistic and internally consistent representation of clinical scheduling demand, enabling systematic evaluation of the proposed scheduling framework under both steady-state and dynamic conditions.

\subsection{LLM-Based Scheduling Framework Development}
Our scheduling framework leverages ChatGPT 5.4 to generate candidate treatment schedules satisfying predefined clinical and operational rules through structured natural language prompts.  The evaluated system should be interpreted as an existing LLM combined with a human-designed prompting framework informed by RT scheduling domain knowledge, rather than as a measure of intrinsic model capability alone. Our strategy is to transform the complex scheduling problem into a structured, executable task for the LLM by dividing the instructions into a \textit{reusable prompt template} and an \textit{instance-specific scheduling prompt},  both provided in the Supplementary Material. Each scheduling task used the same reusable prompt template and a day-specific input prompt. No external code execution, database access, or optimization solver was available to the LLM. The reusable template serves as the foundational logic layer, encoding the overarching scheduling objectives, such as maximizing machine utilization and the  operational constraints. Complementing this, the instance-specific prompt provides the real-time operational context for a given day, including the current state of the treatment calendar, the specific scheduling horizon, and the list of new patients requiring allocation. 

 All scheduling experiments were conducted through the ChatGPT web interface using a standard web browser. As a result, specific inference settings could not be specified such as temperature, top-$p$, random seed, maximum output length, and other generation parameters. Each scheduling request was submitted using the reusable prompt template together with the corresponding instance-specific scheduling input, and the LLM was asked to produce a complete schedule in a single response. Prompt length and context-window limitations were managed by structuring the task as sequential day-by-day scheduling. For each scheduling instance, the prompt included the reusable rules, the active treatment calendar, the scheduling horizon needed for currently active and newly arriving patients, and compact structured patient information. Patients who had completed treatment were removed from the active calendar, which limited unnecessary context growth as the simulation progressed. The generated schedule was not manually edited before validation. 

If an output was incomplete, syntactically unusable, or failed deterministic validation of hard feasibility constraints, it was not accepted as a valid schedule; during prompt-development experiments, such failures motivated the prompt refinements described below. Prompt development and refinement were performed using a separate synthetic scheduling dataset of approximately 40 operating days, which was independent of the 260-day dataset used for the final reported evaluation.

Because LLMs do not inherently know RT scheduling rules or workflow constraints, prompt design was conducted in a staged and structured manner to explicitly encode the logic of the scheduling task. In the initial stage, prompts were constructed to ensure that the LLM correctly understood the fundamental characteristics of RT treatment delivery. Specifically, the model was instructed that most patients undergo multi-fraction treatment courses and, once initiated, treatments must proceed continuously on weekdays (Monday–Friday) until completion. Special treatment patterns, including BID regimens requiring two fractions per day with minimum inter-fraction spacing, were also explicitly described. In the second stage, detailed clinical and operational constraints were introduced. These included machine availability (three LINACs), machine eligibility restrictions for specific patient categories, fixed new-start time windows, extended duration for first-fraction treatments, and category-specific scheduling constraints summarized in Table~\ref{tab:patient_categories}. Explicit references to these tables were incorporated into the prompt to ensure correct interpretation of category-dependent parameters.

Initial deployment of the LLM revealed several systematic failure modes. First, the model occasionally failed to schedule all eligible patients despite available capacity. This arose because the LLM interpreted the task as generating a feasible schedule rather than ensuring complete patient assignment. This issue was resolved by explicitly instructing the model to schedule all patients whenever feasible and to verify that no schedulable patient remained unassigned. Second, early schedules contained unnecessary idle gaps even when machine time was available. While all constraints were satisfied, the LLM did not inherently optimize for compactness. Prompts were refined to explicitly require minimization of unscheduled gaps and compaction of treatments. Third, BID treatments were sometimes scheduled such that the second fraction occurred late in the day, occasionally beyond standard operating hours. Although minimum spacing constraints were satisfied, this behavior was clinically undesirable. The prompt was modified to prioritize early placement of first fractions for multi-fraction regimens, ensuring that subsequent fractions remained within operating hours. Fourth, significant imbalance in workload across LINACs was observed in early outputs. In some cases, one machine completed treatments several hours earlier than others, while remaining machines continued until the end of the clinical day. Although technically feasible, such imbalance is operationally inefficient and inconsistent with clinical practice. To address this, prompts were refined to explicitly require balanced workload distribution, encouraging all LINACs to complete treatments at approximately the same time. Finally, inconsistent enforcement of constraints was observed when rules were stated only once. This issue was mitigated by intentionally repeating critical constraints in multiple parts of the prompt, reinforcing their importance and improving reliability. Through iterative refinement, these modifications defined the final prompt template used in all evaluation experiments.  These refinements represent manual prompt-engineering effort informed by operational failure analysis.

Beyond the reusable prompt template, scenario-specific scheduling behavior was implemented through targeted modifications to the instance-specific scheduling prompt. These additions allowed the same underlying framework to accommodate different operational priorities, including a weekly time-consistency baseline, LINAC continuity, gap-constrained temporal relaxation, and infeasible scheduling request handling. This design allowed the base feasibility rules to remain unchanged while scenario-specific objectives were added or removed, reducing the risk that performance differences were caused by changes to the core rule set.

\subsection{Evaluation}

\subsubsection[Evaluation of the LLM-Based Scheduling Framework]{ Evaluation of the LLM-Based Scheduling Framework}

We evaluated the framework across four scheduling scenarios, which were designed to test whether prompt modifications could incorporate different clinical and operational priorities while preserving core feasibility constraints.

\textbf{Scenario 1: Weekly time-consistency baseline.} In routine clinical practice, patients receiving multi-fraction RT treatments often prefer to receive treatment at approximately the same time each day because consistent treatment times simplify transportation planning, work schedules, caregiver coordination, and daily routines. From the clinic perspective, however, maintaining identical treatment times across all fractions can be difficult because of fluctuating daily workloads, machine availability constraints, and category-specific scheduling requirements. Therefore, the first scenario was designed to evaluate whether the LLM-based framework could maintain clinically practical treatment-time consistency while preserving operational feasibility and machine utilization.

In this scenario, all clinical and operational constraints described in Section~\ref{sec:scheduleconstraints} were enforced together with a temporal consistency objective implemented through a ``weekly one-hour rule.'' Specifically, the difference between the earliest and latest treatment start times within a Monday--Friday treatment week was requested to remain within 60 minutes whenever feasible. The one-hour rule was implemented as a soft scheduling objective rather than a hard constraint. The prompt instructed the LLM to minimize each patient's weekly treatment-time range even when the 60-minute window could not be fully satisfied.

The first treatment fraction for each patient was excluded from the weekly range calculation. The prompt further instructed the model to relax the one-hour preference whenever necessary to satisfy hard clinical constraints, such as BID six-hour separation requirements. In addition, treatment gaps were allowed when necessary to preserve weekly treatment-time consistency, particularly on Fridays when no new patients are initiated and the treatment workload is typically lower. The prompt also included a lower-priority preference encouraging treatment-time consistency across consecutive treatment weeks whenever feasible.

\textbf{Scenario 2: Weekly time consistency with LINAC continuity.}  A machine switch was defined as assignment of two consecutive treatment fractions for the same patient to different LINACs. Although machine switching between treatment days is generally clinically acceptable when machine eligibility constraints are satisfied, both patients and clinical staff often prefer to maintain treatment continuity on the same LINAC whenever possible. Therefore, the second scenario was designed to evaluate whether the LLM framework could incorporate a LINAC-continuity objective while preserving the weekly treatment-time consistency and operational efficiency achieved in Scenario 1.

 As such, an additional LINAC-continuity scheduling objective was introduced to reduce unnecessary machine changes when multiple feasible LINAC assignments were available. While patients remained eligible to receive treatment on multiple LINACs according to Table \ref{tab:patient_categories}, the prompt instructed the LLM to preserve machine continuity whenever feasible, while allowing switches when required to satisfy higher-priority clinical or operational constraints. This objective was added on top of all constraints and scheduling objectives used in Scenario 1.

\textbf{Scenario 3: LINAC continuity with gap-constrained temporal relaxation.}
The first two scenarios demonstrated that strong treatment-time consistency and reduced machine switching could be achieved simultaneously. However, preserving these objectives occasionally introduced substantial idle machine gaps, particularly on Fridays when no new patients are initiated. In real clinical operations, excessive idle gaps may reduce machine utilization efficiency and unnecessarily prolong clinic operating times. Therefore, the third scenario was designed to evaluate whether schedule compactness could be improved by relaxing temporal consistency only when preserving it would create excessive LINAC idle gaps.

In this setting, the prior instruction allowing treatment gaps to preserve weekly treatment-time consistency was removed. Instead, based on practical operational tolerances in our clinic, the prompt instructed the model to relax the treatment-time consistency objective based on a 20-minute threshold defined as an idle interval on an individual LINAC schedule. If preserving a patient’s weekly one-hour window would create an idle interval greater than 20 minutes, the model was allowed to relax the temporal consistency objective.

\textbf{Scenario 4: Infeasible scheduling request handling.}
While the previous scenarios focused on clinically feasible scheduling conditions, real-world RT environments frequently involve unexpected requests or operational disruptions that may render a scheduling request infeasible. Robust handling of such situations is critical. Therefore, the final scenario was designed to evaluate the robustness and interpretability of the framework when confronted with intentionally infeasible scheduling requests.

The experiments were performed by intentionally modifying the simulated patient dataset to introduce infeasible requests and operational conflicts. Ten infeasible scenarios were tested, including assignment of new patients on Fridays, violation of category-specific first-fraction start-day constraints, such as initiating Breast2 patients on non-Mondays, and cases in which the number of new patients exceeded the available first-fraction start slots across the three LINACs. Each scenario included multiple test cases with intentionally introduced infeasible scheduling requirements. These experiments were designed to test the infeasibility-detection and handling logic encoded in the reusable prompt template. Correct behavior was defined as explicit identification of the infeasible request and refusal to silently generate a schedule violating hard constraints.

Table~\ref{tab:scenarioEvaluation} summarizes the four evaluation scenarios, the primary objective of each scenario, the added scheduling rule or test condition, and the corresponding primary metric.
\begin{table}[htbp]
\centering
\caption{Summary of scheduling evaluation scenarios and primary evaluation metrics.}
\label{tab:scenarioEvaluation}
\begin{tabularx}{\textwidth}{cXXX}
\hline
Scenario & Main objective & Added rule & Primary metric \\
\hline
1 & Weekly time consistency 
  & 60-min soft window 
  & Percentage of fractions within 60 min \\

2 & LINAC continuity 
  & Reduce unnecessary LINAC switching 
  & Switch rate \\

3 & Schedule compactness 
  & Relax the weekly time window when preserving it would create LINAC idle gaps greater than 20 min 
  & Friday gap \\

4 & Infeasible request handling 
  & Test intentionally invalid scheduling requests 
  & Correct detection and reporting of scheduling conflicts \\
\hline
\end{tabularx}
\end{table}

\textbf{Evaluation}
 After schedules were generated, the feasibility was evaluated using a deterministic rule-based validation program implemented in Python. The same program was applied consistently across scenarios. The program explicitly checked complete patient assignment, treatment duration, fraction sequencing, absence of overlapping treatments on the same LINAC, machine eligibility, BID spacing, first-fraction start windows, anesthesia timing constraints, and overtime. In addition, randomly selected schedules and patient cases were manually reviewed as an independent spot-checking procedure. Feasibility was considered as confirmed only when both deterministic validation and manual review identified no violations.

 After feasibility was confirmed, schedule quality was quantified using seven clinically and operationally meaningful metrics. Mean gap measured unused time between consecutive appointments across the clinic schedule and reflected schedule compactness and available treatment capacity. Friday gap was reported separately because no new-start patients were scheduled on Fridays, making Friday idle time a sensitive measure of end-of-week schedule compactness. LINAC switch rate quantified the percentage of evaluable consecutive-fraction transitions assigned to a different LINAC and reflected treatment-course continuity and operational simplicity. Time consistency rate quantified the percentage of evaluated fractions satisfying the preferred $\leq$60-min weekly treatment-time window. Overtime (OT) burden was described using three complementary measures: OT rate, daily OT, and max OT. These overtime metrics reflect the extent to which a schedule extends beyond the standard operating day and therefore relate to staffing burden, patient flow, and operational feasibility.

 To estimate both overall performance and robustness of the scheduling approach over a relatively long period, the 260 operating days were divided into 13 consecutive four-week segments, each containing 20 operating days. Within each segment, mean gap, Friday gap, LINAC switch rate, and time consistency rate were calculated; OT rate was computed as the percentage of operating days with at least one appointment finishing after 5:00 PM; and daily OT was calculated as the average overtime minutes beyond 5:00 PM per day. These metrics were summarized as mean $\pm$ standard deviation across the 13 segments. Max OT was reported as the maximum overtime observed over the full 260-day evaluation period, without a segment-level standard deviation.

\subsubsection[Optimization-Based Scheduling Approaches and Comparisons]{ Optimization-Based Scheduling Approaches and Comparisons}

 To provide a non-LLM benchmark for interpreting the LLM-generated schedules, we implemented optimization-based scheduling approaches using the same 260-day simulated scheduling dataset. The benchmark was intended to provide a preliminary reference for evaluating LLM-generated candidate schedules under matched simulated conditions, rather than to establish superiority over conventional optimization-based scheduling. Here, we briefly present the optimization model and algorithm due to space limitation. For details, please refer to the Supplementary Material. 

 Let $A$ denote the set of all appointments to be scheduled, with $a \in A$ indexing an individual appointment, and let $N\subseteq A$ denote the subset of first-fraction appointments for new-start patients. Let $M$ denote the set of LINACs and $m \in M$ an individual LINAC. The set $T$ denotes the discrete candidate start-time grid, which used a 5-minute resolution to accommodate treatment durations such as the 25-minute multi-isocenter category. For each appointment $a$, $E_a \subseteq M$ is the set of eligible LINACs, and $\bar{p}_a$ is the base treatment duration. Let $I_a$ indicate whether appointment $a$ is a first fraction and let $u$ denote the additional first-fraction setup time. The appointment duration was therefore defined as $p_a=\bar{p}_a+uI_a$, using the same duration rules as in Section~\ref{sec:scheduleconstraints}. The variables $s_a$ and $e_a$ denote the scheduled start and end times, and $x_{a,m,t}$ is a binary assignment variable equal to 1 if appointment $a$ is assigned to LINAC $m$ at start time $t$. The assignment variable $x_{a,m,t}$ is the primary decision variable; $s_a$, $e_a$, $s_{k,d,m}$, $e_{k,d,m}$, and $f_{d,m}$ are derived schedule quantities determined by the selected assignments, where $k$ denotes the within-day order of appointments assigned to LINAC $m$ on operating day $d$. Let $B$ denote the set of ordered same-day BID pairs, where $(b_1,b_2)\in B$ represents the first and second fractions for the same patient on the same day, and let $g=360$ minutes denote the required minimum spacing between BID fractions. The value $h_a$ denotes the prior or preferred treatment time for appointment $a$.

 Each appointment was assigned exactly once to an eligible LINAC and feasible start time. Hence, $\sum_{m \in E_a}\sum_{t \in T} x_{a,m,t}=1,
~x_{a,m,t} \in \{0,1\}, ~\forall a \in A$.
The scheduled start and end times were defined as $s_a=\sum_{m \in E_a}\sum_{t \in T} t\,x_{a,m,t},~e_a=s_a+p_a$.
No two appointments were allowed to overlap on the same LINAC, so $\sum_{a \in A}\sum_{t \in T:\,\tau \in [t,t+p_a)}x_{a,m,t} \le 1,~ \forall m \in M, \; \forall \tau \in T$.
For each ordered BID pair $(b_1,b_2)\in B$, the second fraction was required to start at least $g$ minutes after the first fraction. Therefore, $s_{b_2} \ge s_{b_1}+g$. Additional hard constraints enforced LINAC eligibility, valid first-fraction start days, allowed first-fraction start-time windows, standard operating hours, and same-day BID spacing. Only candidate assignments satisfying these feasibility rules were considered in the optimization.

 The multi-objective scheduling cost $Z$ was written as
\begin{equation}
Z =
w_GG+w_SS+w_DD+w_QQ+w_LL+w_OO+w_FF+w_CC.
\end{equation}
Here, $G$ penalizes idle gaps between consecutive appointments on the same LINAC and day; $S$ penalizes imbalance in daily LINAC finish times; $D$ penalizes deviation from a prior or preferred treatment time; $Q$ penalizes inter-fraction LINAC switching; $L$ penalizes late-day placement after 4:00 PM; $O$ penalizes overtime beyond 5:00 PM; $F$ penalizes the latest daily finish time; and $C$ is an end-time compactness surrogate used in the CP-SAT implementation\cite{perron2023cp}. The optimization benchmark was evaluated using two objective-weight profiles: a balanced profile emphasizing treatment-time consistency and LINAC continuity, and a finish-time-priority profile emphasizing compact daily schedules, reduced idle gaps, and reduced overtime. Detailed objective weights and implementation settings are provided in the Supplementary Material.

 We implemented two solvers for this optimization problem. The first was a daily-decomposition heuristic solver using sequential insertion. For each operating day, appointments were ordered by scheduling priority, including new-start BID pairs, other new starts, continuing BID first fractions, and remaining fractions. For each appointment, the heuristic enumerated feasible LINAC and start-time candidates satisfying hard constraints, scored each feasible candidate using the weighted objective terms, inserted the lowest-cost candidate, and then proceeded to the next appointment. Same-day BID treatments were handled as paired first- and second-fraction decisions to enforce the six-hour spacing requirement.

 The second solver used a constraint-programming formulation implemented with OR-Tools CP-SAT\cite{perron2023cp}. Each daily scheduling problem was formulated with interval variables and machine-capacity constraints, with no-overlap constraints enforcing that each LINAC could treat at most one appointment at a time. Linear constraints enforced appointment assignment, machine eligibility, operating-hour rules, first-fraction requirements, and BID spacing. CP-SAT was implemented as a time-limited daily optimization benchmark; when optimality was not proven within the solver limit, the best feasible solution found by the solver was retained. Therefore, the reported CP-SAT schedules should be interpreted as feasible solver-generated benchmark schedules rather than necessarily proven globally optimal schedules.

{\revision Note that these implementations served as illustrative computational benchmarks and were not exhaustively tuned; CP-SAT optimality was not systematically assessed. Different objective-weight profiles were used to illustrate how scheduling preferences affect performance tradeoffs, without claiming superiority over other optimization formulations or tuning choices.}

The resulting optimization schedules were validated with the same deterministic rule-based validation program used for the LLM-generated schedules. Performance metrics were then calculated and compared with LLM-generated schedules.

\subsubsection[Retrospective Comparison with Clinical Scheduling]{\revision Retrospective Comparison with Clinical Scheduling}
\label{sec:clinicalcomparison}

{\revision We further conducted a six-week retrospective test using real clinical scheduling data. Specifically, the evaluation covered 30 operating days (six weeks) across three LINACs and 2,043 appointments, including 898 appointments existing at the beginning of this six-week period for patients on treatment, and 1,145 appointments from 141 new patients to schedule. All seven approaches (three LLM approaches and four optimization approaches) used the same initial schedule and new-patient workload. New patients entered sequentially by required start day and arrival order following the actual clinical scenarios. The resulting schedules were evaluated over the same period using the metric definitions previously.}

\section{Results}

\subsection[LLM Scheduling Results]{ LLM Scheduling Results}

\subsubsection[Representative LLM Results]{ Representative LLM Results}

\begin{figure}[p]
    \centering
    \includegraphics[width=\textwidth,height=0.88\textheight,keepaspectratio]{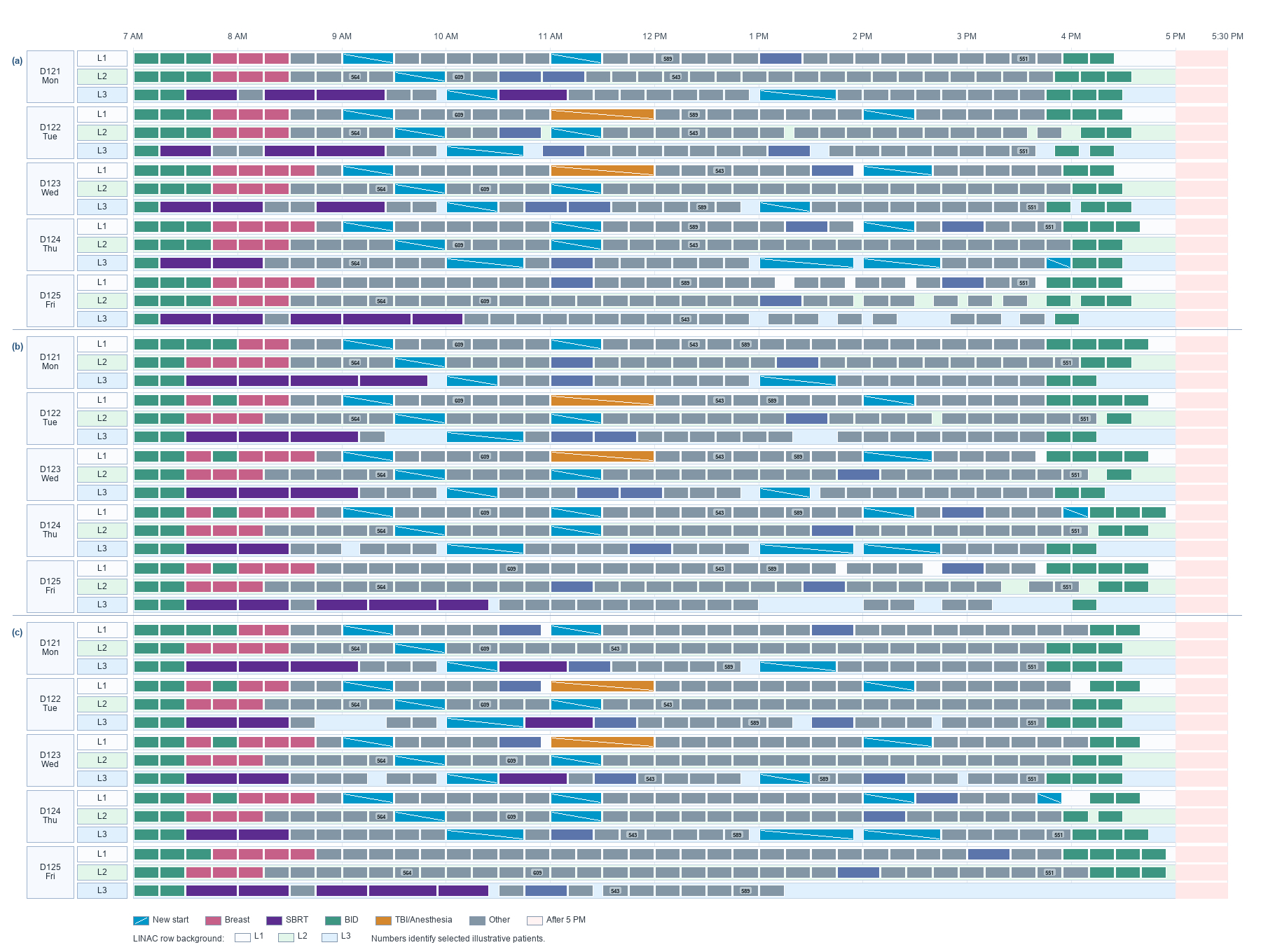}
    \caption{(a)-(c) Representative LLM-generated schedules for days 121--125 under the three evaluated scheduling scenarios. Each day is displayed as a separate block, with the three LINACs L1, L2, and L3 rows shown using distinct background colors. Numeric labels identify selected multi-day patients to illustrate treatment-time consistency and LINAC assignment patterns across the representative week.}
    \label{fig:representative_ai_schedules}
\end{figure}

 Across the evaluated scheduling scenarios, the LLM-based framework successfully generated complete candidate schedules. The representative schedules demonstrate that the model outputs were not limited to isolated single-day examples, but could organize multi-day treatment courses across three LINACs while respecting the structured scheduling rules provided in the prompts. Representative schedules for scenarios 1--3 are shown in Fig.~\ref{fig:representative_ai_schedules}. In the weekly time-consistency baseline, repeated fractions for the same patient were generally placed at similar treatment times across the week. After adding the LINAC-continuity objective, the same patient was more consistently assigned to the same machine across treatment days. In the gap-constrained temporal-relaxation scenario, the schedules preserved most of this temporal and machine consistency while allowing modest placement flexibility to reduce idle gaps, particularly on lower-volume days. Overall, these representative results indicate that the prompt-guided LLM scheduler produced clinically interpretable schedules that were broadly aligned with the predefined operational constraints and scenario-specific scheduling objectives.

\subsubsection[Scenario-Based LLM Scheduling Performance]{ Scenario-Based LLM Scheduling Performance}
\label{sec:ai_scenario_performance}
\label{sec:1h_baseline}
\label{sec:1h_reducedswitching}

 The three LLM scheduling scenarios evaluated progressively more complex operational preferences and were compared using the seven predefined performance metrics summarized in Table~\ref{tab:performance_summary}.

\begin{table}[htbp]
\centering
\caption{Summary comparison of LLM scheduling performance across the three feasible scheduling scenarios. Values are reported as mean $\pm$ standard deviation across 13 consecutive four-week evaluation segments, except maximum OT, which is reported over the full 260-day evaluation period.}
\label{tab:performance_summary}
\footnotesize
\setlength{\tabcolsep}{1.5pt}
\renewcommand{\arraystretch}{1.15}

\begin{tabular}{
>{\centering\arraybackslash}m{0.82in}
>{\centering\arraybackslash}p{0.76in}
>{\centering\arraybackslash}p{0.78in}
>{\centering\arraybackslash}p{0.78in}
>{\centering\arraybackslash}p{0.92in}
>{\centering\arraybackslash}p{0.64in}
>{\centering\arraybackslash}p{0.76in}
>{\centering\arraybackslash}p{0.62in}
}
\hline
\makecell{Model/\\Scenario} &
\makecell{Mean\\gap\\(min)} &
\makecell{Friday\\gap\\(min)} &
\makecell{LINAC\\switch\\rate\\(\%)} &
\makecell{Time\\consistency\\rate\\(\%)} &
\makecell{OT rate\\(\%)} &
\makecell{Daily\\OT\\(min)} &
\makecell{Max\\OT\\(min)} \\
\hline
LLM S1 &
57.4 $\pm$ 9.2 &
159.6 $\pm$ 18.7 &
52.4 $\pm$ 2.6 &
98.7 $\pm$ 1.0 &
1.2 $\pm$ 3.0 &
0.1 $\pm$ 0.2 &
15 \\

LLM S2 &
62.8 $\pm$ 9.5 &
169.5 $\pm$ 19.5 &
9.4 $\pm$ 0.6 &
98.8 $\pm$ 1.0 &
0.8 $\pm$ 1.9 &
0.1 $\pm$ 0.2 &
15 \\

LLM S3 &
47.1 $\pm$ 8.5 &
89.2 $\pm$ 18.4 &
10.5 $\pm$ 1.4 &
98.6 $\pm$ 1.0 &
1.2 $\pm$ 3.0 &
0.1 $\pm$ 0.3 &
15 \\
\hline
\end{tabular}

\end{table}

Scenario~1 served as the weekly treatment-time consistency baseline. Across the 13 four-week evaluation segments, mean gap was 57.4 $\pm$ 9.2 minutes, and Friday gap was higher at 159.6 $\pm$ 18.7 minutes, consistent with the no-Friday-new-start rule. Time consistency rate was strong, with 98.7 $\pm$ 1.0\% of evaluated fractions satisfying the preferred 60-minute weekly window. This aggregate metric was consistent with the detailed distribution in the Supplementary Material: 15,345 of 25,137 evaluated fractions (61.0\%) had a weekly treatment-time difference of 0 minutes, and 24,837 fractions (98.8\%) were within 60 minutes over the full 260-day evaluation period. 
Fractions exceeding the 60-minute threshold tended to occur on days containing TBI or anesthesia cases, for which rigid treatment durations and timing requirements reduced scheduling flexibility for other patients.

 Scenario~2 tested whether a LINAC-continuity objective could be added on top of the weekly time-consistency baseline. LINAC switch rate decreased from 52.4 $\pm$ 2.6\% in Scenario~1 to 9.4 $\pm$ 0.6\% after adding the continuity objective, while time consistency rate remained high at 98.8 $\pm$ 1.0\%. The detailed full-period analysis in the Supplementary Material similarly showed a reduction from 11,403 switches under Scenario~1 to 2,103 switches under Scenario~2 among 20,893 evaluable consecutive-fraction transitions. The trade-off was a modest increase in idle gaps, with mean gap increasing to 62.8 $\pm$ 9.5 minutes and Friday gap increasing to 169.5 $\pm$ 19.5 minutes.

 Scenario~3 then incorporated gap-constrained temporal relaxation to improve schedule compactness while preserving the main benefits of time consistency and LINAC continuity. This scenario produced the lowest mean gap among the scenarios at 47.1 $\pm$ 8.5 minutes and reduced Friday gap to 89.2 $\pm$ 18.4 minutes. The detailed weekday analysis in the Supplementary Material showed the same pattern, with the Friday median gap decreasing from 165.0 minutes in Scenario~2 to 80.0 minutes in Scenario~3, although the maximum Friday gap remained 280 minutes. LINAC continuity was largely preserved, with a LINAC switch rate of 10.5 $\pm$ 1.4\%, and time consistency rate remained high at 98.6 $\pm$ 1.0\%. In the full-period distribution, 24,846 of 25,137 evaluated fractions (98.8\%) remained within the preferred 60-minute weekly window, indicating that the relaxation mainly affected a small number of constrained cases.

 Across the three scenarios, overtime remained rare. The OT rate was 1.2 $\pm$ 3.0\%, 0.8 $\pm$ 1.9\%, and 1.2 $\pm$ 3.0\% for scenarios 1--3, respectively, with daily OT of 0.1 minutes on average in each scenario and a maximum OT duration of 15 minutes. Detailed review showed that these overtime cases occurred under near-capacity workload conditions and often involved BID patients receiving their second daily fraction near the end of the day.

\subsubsection[Infeasible Request Handling]{ Infeasible Request Handling}

Across all ten infeasible scheduling request test scenarios, the LLM consistently identified the scheduling conflicts and responded according to the predefined infeasibility-handling instructions embedded in the reusable prompt template. The model correctly detected prohibited first-fraction start days, category-specific start-day violations, and cases in which the number of new patients exceeded the available first-fraction start slots across the three LINACs.

Rather than generating invalid schedules, the LLM explicitly identified the source of infeasibility and proposed actionable alternatives, such as moving patients to the next valid start day or requesting human review when the requested schedule conflicted with predefined rules. No cases were observed in which the model silently violated clinical or operational rules while attempting to force a feasible schedule.

These findings indicate that the proposed framework is capable not only of generating candidate treatment schedules satisfying predefined clinical and operational rules, but also of exhibiting robust and interpretable behavior when confronted with infeasible scheduling requests. 

\subsection[Comparison with Optimization-based Approaches]{ Comparison with Optimization-based Approaches}

 The optimization-based benchmarks generated complete schedules for the same 260-day simulated case set used in the LLM experiments. All optimization schedules satisfied the hard feasibility rules. The optimization results are summarized in Table~\ref{tab:optimization_performance} using the same metrics reported for the LLM scheduling in Table~\ref{tab:performance_summary}. Representative weekly schedules for the four optimization-based schedules are provided in the Supplementary Material.

\begin{table}[htbp]
\centering
\caption{Summary comparison of optimization-based scheduling performance. Values are reported as mean $\pm$ standard deviation across 13 consecutive four-week evaluation segments, except maximum OT, which is reported over the full 260-day evaluation period.}
\label{tab:optimization_performance}
\footnotesize
\setlength{\tabcolsep}{1.5pt}
\renewcommand{\arraystretch}{1.15}

\begin{tabular}{
>{\centering\arraybackslash}m{0.82in}
>{\centering\arraybackslash}p{0.76in}
>{\centering\arraybackslash}p{0.78in}
>{\centering\arraybackslash}p{0.78in}
>{\centering\arraybackslash}p{0.92in}
>{\centering\arraybackslash}p{0.64in}
>{\centering\arraybackslash}p{0.76in}
>{\centering\arraybackslash}p{0.62in}
}
\hline
\makecell{Model/\\Scenario} &
\makecell{Mean\\gap\\(min)} &
\makecell{Friday\\gap\\(min)} &
\makecell{LINAC\\switch\\rate\\(\%)} &
\makecell{Time\\consistency\\rate\\(\%)} &
\makecell{OT rate\\(\%)} &
\makecell{Daily\\OT\\(min)} &
\makecell{Max\\OT\\(min)} \\
\hline
\makecell{Heur.\\balanced} &
122.6 $\pm$ 14.8 &
242.7 $\pm$ 32.0 &
24.3 $\pm$ 2.7 &
94.3 $\pm$ 0.9 &
15.8 $\pm$ 7.6 &
4.0 $\pm$ 2.9 &
30 \\

\makecell{Heur.\\finish time} &
51.6 $\pm$ 7.1 &
42.0 $\pm$ 19.2 &
41.0 $\pm$ 3.2 &
73.4 $\pm$ 1.8 &
5.0 $\pm$ 5.4 &
1.0 $\pm$ 1.5 &
15 \\

\makecell{CP-SAT\\balanced} &
93.9 $\pm$ 21.7 &
202.4 $\pm$ 41.9 &
4.6 $\pm$ 0.9 &
90.0 $\pm$ 2.2 &
29.2 $\pm$ 11.5 &
4.8 $\pm$ 2.1 &
15 \\

\makecell{CP-SAT\\finish time} &
58.1 $\pm$ 13.3 &
112.6 $\pm$ 51.1 &
21.1 $\pm$ 5.3 &
45.2 $\pm$ 5.2 &
20.0 $\pm$ 11.5 &
3.2 $\pm$ 2.1 &
15 \\
\hline
\end{tabular}

\end{table}

 The optimization benchmarks demonstrated that conventional optimization could produce feasible schedules under the same simulated conditions, but the resulting trade-offs depended strongly on the solver and objective-weight profile. The heuristic finish-time profile reduced idle gaps and overtime burden relative to the heuristic balanced profile, including a lower Friday gap, lower OT rate, and lower daily OT, but this improvement was accompanied by a higher LINAC switch rate and reduced time consistency rate. The CP-SAT balanced profile achieved the lowest LINAC switch rate among the optimization benchmarks, but had larger idle gaps, a higher OT rate, greater daily OT, and lower time consistency rate than the LLM-generated schedules. The CP-SAT finish-time profile reduced mean gap relative to CP-SAT balanced scheduling, but substantially reduced time consistency rate.

 Compared with the LLM scheduling results, the optimization results highlight the expected trade-off between explicit objective weighting and balanced clinical performance. In this preliminary comparison, the LLM S3 schedule in Table~\ref{tab:performance_summary} had the lowest mean gap, maintained a low LINAC switch rate, preserved approximately 99\% of evaluated fractions within the preferred 60-minute weekly window, and produced only limited overtime. The heuristic finish-time optimization achieved a lower Friday gap than LLM S3, but at the cost of substantially higher LINAC switch rate and lower time consistency rate. Conversely, CP-SAT balanced optimization achieved lower LINAC switch rate than LLM S3, but with higher idle gaps, more overtime, and lower time consistency rate.

\begin{figure}[htbp]
\centering
\includegraphics[width=\textwidth,height=0.88\textheight,keepaspectratio]{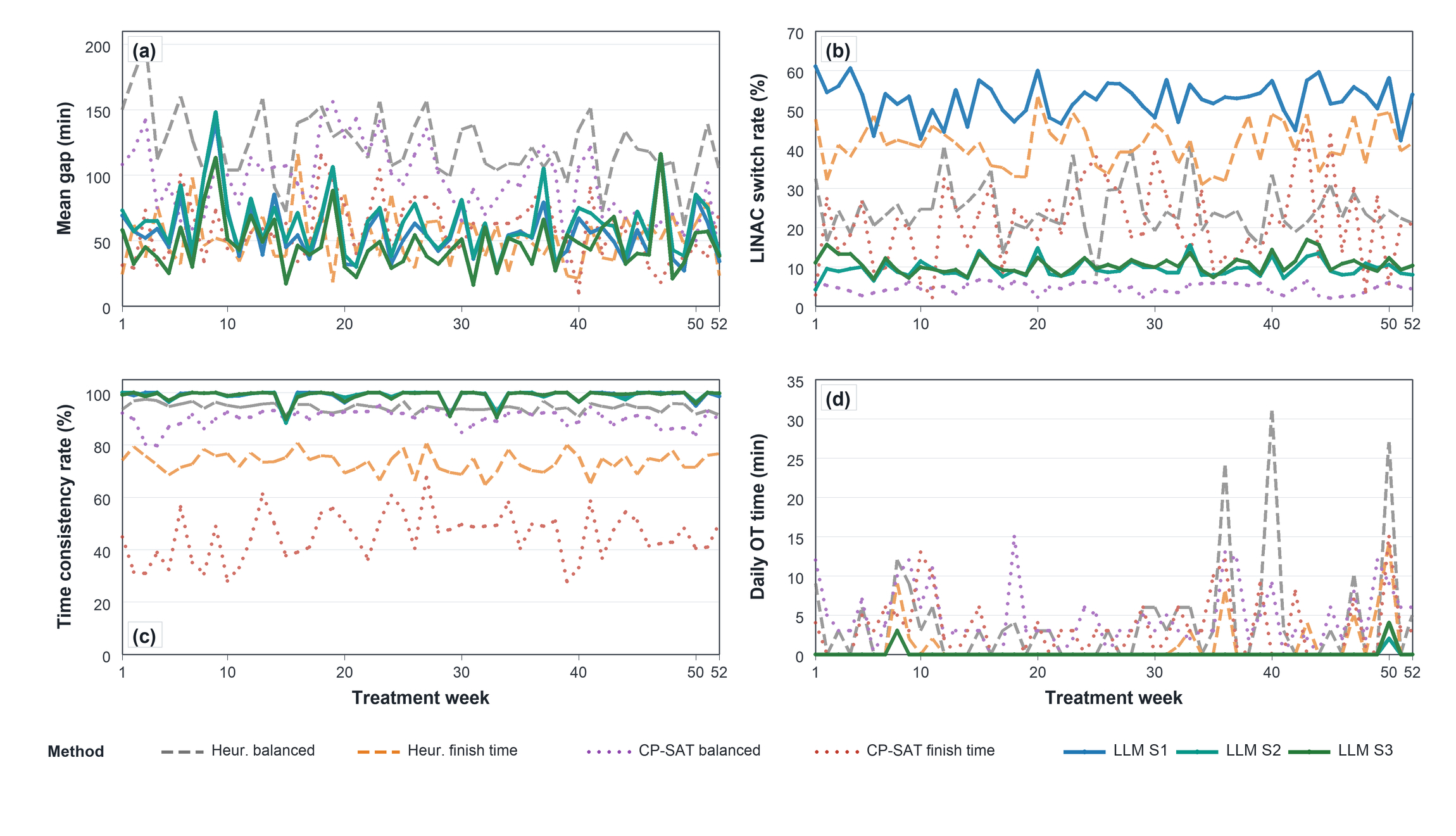}
\caption{Weekly comparison of LLM and optimization-based scheduling performance across the 260-day simulated case set. Each point summarizes one consecutive 5-day treatment week for all seven evaluated methods: (a) Mean gap, (b) LINAC switch rate, (c) Time consistency rate, and (d) Daily OT.}
\label{fig:weekly_performance_comparison}
\end{figure}

 The week-by-week comparison in Fig.~\ref{fig:weekly_performance_comparison} provides a visualization of these trade-offs across the 260-day simulated period. The LLM approach maintained high time consistency rate across nearly all treatment weeks, while the LINAC-continuity scenarios, LLM S2 and LLM S3, showed consistently lower LINAC switch rate than LLM S1. Across the four optimization benchmarks, lower idle gap or lower switching performance was generally accompanied by reduced time consistency rate or increased overtime burden, depending on the objective-weight profile. We would like to remark that these findings do not establish global optimality, superiority over classical scheduling methods, or readiness for clinical replacement; rather, they indicate that the prompt-guided LLM framework produced a feasible and clinically interpretable compromise across the evaluated scheduling objectives.

\subsection[Comparison with Clinical Scheduling Data]{\revision Comparison with Clinical Scheduling Data}
\label{sec:clinicalresults}

{\revision Table~\ref{tab:clinical_performance} summarizes the clinical schedule and all seven computational approaches.All seven computational approaches scheduled every appointment in the evaluation window without overtime, whereas the clinical schedule had overtime on 20.0\% of days. The LLM strategies reduced mean gap by 50.8--77.1\% and Friday gap by 42.7--80.3\% relative to the clinical schedule, while increasing time consistency rate from 86.7\% to 99.5--99.9\%. These improvements are consistent with the greater scheduling flexibility available at routine workload, while the differences among strategies reflect their scheduling objectives.}

\begin{table}[htbp]
\centering
\revision
\caption{Scheduling performance over six weeks.}
\label{tab:clinical_performance}
\footnotesize
\setlength{\tabcolsep}{1.5pt}
\renewcommand{\arraystretch}{1.15}
\begin{tabular}{
>{\centering\arraybackslash}m{0.82in}
>{\centering\arraybackslash}p{0.76in}
>{\centering\arraybackslash}p{0.78in}
>{\centering\arraybackslash}p{0.78in}
>{\centering\arraybackslash}p{0.92in}
>{\centering\arraybackslash}p{0.64in}
>{\centering\arraybackslash}p{0.76in}
>{\centering\arraybackslash}p{0.62in}
}
\hline
\makecell{Method} &
\makecell{Mean\\gap\\(min)} &
\makecell{Friday\\gap\\(min)} &
\makecell{LINAC\\switch\\rate\\(\%)} &
\makecell{Time\\consistency\\rate\\(\%)} &
\makecell{OT rate\\(\%)} &
\makecell{Daily\\OT\\(min)} &
\makecell{Max\\OT\\(min)} \\
\hline
\makecell{Clinical\\schedule} & 466.2 & 567.5 & 5.6 & 86.7 & 20.0 & 6.2 & 60.0 \\
LLM S1 & 106.8 & 186.7 & 52.6 & 99.9 & 0.0 & 0.0 & 0.0 \\
LLM S2 & 229.5 & 325.0 & 1.3 & 99.5 & 0.0 & 0.0 & 0.0 \\
LLM S3 & 130.3 & 111.7 & 7.7 & 99.5 & 0.0 & 0.0 & 0.0 \\
\makecell{Heur.\\balanced} & 218.8 & 243.3 & 27.2 & 97.8 & 0.0 & 0.0 & 0.0 \\
\makecell{Heur.\\finish time} & 18.2 & 3.3 & 47.2 & 63.7 & 0.0 & 0.0 & 0.0 \\
\makecell{CP-SAT\\balanced} & 379.3 & 510.0 & 9.3 & 97.5 & 0.0 & 0.0 & 0.0 \\
\makecell{CP-SAT\\finish time} & 178.8 & 146.7 & 27.5 & 68.4 & 0.0 & 0.0 & 0.0 \\
\hline
\end{tabular}
\end{table}

{\revision LLM S1 emphasizes weekly treatment-time consistency and achieved the highest time consistency rate, 99.9\%, with a mean gap of 106.8 minutes. Its LINAC switch rate of 52.6\% indicates a substantial tradeoff in machine continuity when preferred treatment times are prioritized. S2 adds a LINAC-continuity objective and reduced LINAC switch rate to 1.3\%, while retaining a time consistency rate of 99.5\%. Its larger mean gap of 229.5 minutes and Friday gap of 325.0 minutes are consistent with the reduced flexibility associated with keeping consecutive fractions on the same machine. S3 reduced mean gap to 130.3 minutes and Friday gap to 111.7 minutes, while preserving a time consistency rate of 99.5\% and a relatively low LINAC switch rate of 7.7\%. Together, these results show how the LLM strategies respond to the scheduling objectives.}

{\revision In this retrospective comparison, the clinical schedule had a low LINAC switch rate of 5.6\%, which may reflect the human scheduler's strong emphasis on machine continuity, alongside relatively larger gaps, lower time consistency, and some overtime. Differences between the clinical and LLM-generated schedules may also reflect additional workflow constraints in clinical practice that were not fully captured in this evaluation.}

{\revision The optimization approaches also improved compactness and eliminated overtime relative to the clinical schedule. The balanced methods increased time consistency to 97.5--97.8\%, although switching rose to 9.3--27.2\%. The finish-time methods achieved smaller gaps at the expense of time consistency and machine continuity.}

\section{Discussion}

\subsection{Study Scope}

 This study evaluated the feasibility of using an existing general-purpose LLM as a reasoning and heuristic scheduling aid for RT scheduling. To enable this evaluation, we developed a structured prompting framework that organizes clinical rules, operational constraints, current schedule states, and scenario-specific objectives into reusable and instance-specific prompts. The central question was whether, when guided by this framework, the LLM could translate complex scheduling instructions into candidate schedules satisfying predefined feasibility criteria under realistic simulated conditions. Future comparative studies are needed to extend this initial benchmark by further quantifying optimality, performance across various settings, and clinical workflow benefit. {\revision The increased-workload simulation and the retrospective clinical comparison address complementary questions. The former examines feasibility and competing objectives in a more demanding scheduling environment, while the latter evaluates scheduling performance relative to a clinical schedule at routine workload.}

 The present study addressed the scheduling task for routine regular clinical operations, but not urgent schedule adjustment. Although Scenario 4 evaluated whether the framework could identify infeasible requests and avoid silently generating schedules that violated hard constraints, the simulation did not model urgent add-ons, same-day sim-and-treat cases, emergency replanning, or machine downtime. The LINAC-continuity objective was intended to discourage unnecessary switching during routine scheduling, not to prevent clinically necessary reassignment when a machine becomes unavailable; in such cases, machine availability and patient eligibility would represent higher-priority constraints that override the continuity preference. In routine practice, same-day sim-and-treat cases are often inserted late in the day because earlier appointment slots are already occupied. Such late-day insertions would likely affect multiple evaluation metrics. In addition, creating room for an urgent patient or recovering from machine downtime may require therapists or scheduling coordinators to identify movable appointments, reassign affected patients to equivalent eligible machines, contact patients, negotiate later or alternative treatment times, and confirm patient acceptance. These communication and patient-preference workflows were outside the scope of the present simulation and cannot be replaced by the LLM framework evaluated here.

\subsection{Computational Considerations and Scheduling Complexity}

 The RT patient scheduling problem is inherently complex, involving a large number of interdependent clinical and operational constraints. In the present benchmark, however, daily decomposition made each optimization subproblem computationally tractable, and the heuristic and CP-SAT implementations generated feasible daily schedules rapidly under the selected time limits. Therefore, the practical contrast between the LLM framework and optimization-based scheduling in this study should not be interpreted primarily as a runtime advantage for the LLM. Instead, the key difference lies in how scheduling logic is specified and modified. Optimization-based approaches require objectives, soft constraints, priorities, and exception-handling rules to be explicitly formulated, weighted, implemented, and maintained. By contrast, the LLM-based framework generates schedules through prompt-guided reasoning over structured clinical rules and natural-language operational objectives, shifting part of the practical burden from mathematical reformulation and parameter tuning toward prompt specification, validation, and human review.

Generating an updated treatment schedule required approximately five minutes. This runtime was measured from prompt submission to complete schedule output and did not include post hoc validation.  This response time did not impede the retrospective feasibility experiments, but it would be an important practical consideration for real-time clinical use, particularly for patient call-in requests that require immediate schedule adjustment.  However, prospective real-time deployment would likely require further optimization of response time, integration with schedule-management systems, and rapid post hoc validation before candidate changes could be discussed with patients.  Thus, the LLM framework should be viewed as a prompt-guided heuristic scheduling aid whose potential value is interpretability and flexible objective management in a human-in-the-loop workflow, rather than faster computation than an appropriately formulated optimization solver.

 Model-version behavior was considered as part of the practical computational characteristics of the proposed framework. This project was initiated using ChatGPT 5.2. As newer versions became available during
the development process, including ChatGPT 5.3 and ChatGPT 5.4, the scheduling experiments
were repeated using the updated models. The final results reported in this study are based on ChatGPT 5.4. Across the ChatGPT versions tested during development, 5.4 generally produced more stable responses. In contrast, 5.2 and 5.3 occasionally stalled for extended periods without producing a complete schedule.

\subsection[LLM-Based Objective Management in Human-in-the-Loop Scheduling]{ LLM-Based Objective Management in Human-in-the-Loop Scheduling}

 The practical meaning of the reported operational metrics should be interpreted in this hierarchy. Complete and feasible treatment delivery on the planned treatment day remains the primary requirement. The weekly one-hour rule was used as a soft operational preference rather than a hard constraint, providing a practical balance between patient-centered treatment-time consistency and scheduling flexibility under competing constraints. A high LINAC switch rate may still be acceptable if all treatments are delivered safely and on schedule; however, unnecessary switching can increase therapist workload, complicate patient flow, and create avoidable operational complexity. Similarly, Friday gap reflects idle machine time and schedule compactness on a day when no new patients are initiated and workload is often lower. Reducing this gap may improve machine utilization, reduce inefficient staff time, and preserve capacity for late-day operational needs, but it remains a soft operational objective rather than a primary clinical endpoint.

 The comparison with optimization-based scheduling results provided an initial benchmark for interpreting the LLM-generated schedules. Accordingly, these results should be interpreted as evidence supporting the feasibility of structured LLM-assisted schedule generation within an operational decision-support workflow, rather than as evidence that LLMs are superior to, ready to replace, or should be deployed in place of mathematical optimization methods or expert clinical schedulers. Based on these results, the main potential value of LLM-based scheduling is a flexible natural-language interface for generating schedules in a human-in-the-loop decision-support workflow. RT patient scheduling requires simultaneous consideration of multiple competing clinical and operational goals. Conventional optimization approaches can address these multi-objective trade-offs, but the model is constrained by the terms that are explicitly represented in the objective function and constraints. Adjusting objective weights can steer the resulting schedule toward different operational priorities, as shown in the optimization comparison, but this process can be technically complex, formulation-dependent, and difficult for clinical users to modify without certain optimization expertise.

 In this context, a potential advantage of the LLM-based framework is that scheduling objectives can be introduced, modified, or reprioritized through structured natural-language instructions. In the present study, the same prompting framework was adapted across different scenarios by adding or modifying objectives, while preserving the same hard feasibility rules. This was reflected in the scenario comparisons: adding the LINAC-continuity objective reduced LINAC switch rate from 52.4 $\pm$ 2.6\% to 9.4 $\pm$ 0.6\% but modestly increased mean gap, whereas gap-constrained temporal relaxation reduced Friday gap from 169.5 $\pm$ 19.5 minutes to 89.2 $\pm$ 18.4 minutes with only a small increase in LINAC switch rate. This flexibility is particularly relevant for clinical scheduling, where local priorities, operational preferences, and patient accommodations may change over time. In a human-in-the-loop workflow, an expert scheduler could express these priorities in natural language, review the resulting candidate schedule, and iteratively request targeted revisions. Thus, the LLM should be viewed as an interactive scheduling aid that may complement other scheduling methods by helping translate nuanced and evolving clinical intent into feasible candidate schedules.

\subsection[Scalability, Reproducibility, and Multi-Institution Applicability] {Scalability, Reproducibility, and Multi-Institution Applicability}

An important potential advantage of the proposed LLM-based framework is its adaptability to diverse clinical environments. This suggests that the approach may be extendable to larger clinical settings or adaptable across institutions.  Modeling the workflow after our three-LINAC department provided realism with respect to our clinic, allowing the simulation to include practical constraints. The trade-off is that some rules reflect institution-specific practice and may not transfer directly to clinics with different machine configurations, staffing models, treatment techniques, or scheduling policies. Future implementations should therefore separate generalizable scheduling logic from locally configurable components, including prompt templates, rule validators, machine-eligibility tables, operating hours, first-fraction policies, and institutional scheduling preferences. Multi-institutional validation will be necessary to determine which components can be standardized and which require local adaptation.

 Scalability to larger facilities also requires direct evaluation. A clinic with a large number of LINACs would increase the number of possible machine-time assignments, machine-eligibility combinations, workload-balancing choices, and interactions among competing constraints. Although the prompt structure can in principle be expanded to include additional machines and local rules, the present three-LINAC simulation does not establish that schedule quality, runtime, or output completeness would remain stable at larger institutional scales.

 At the same time, the prompt-refinement process and the web-based LLM interface are important reproducibility considerations. The final reusable prompt was the result of manual, domain-informed iteration rather than a standardized automated optimization procedure. As a result, the exact refinement process may not transfer directly to different settings, although the principle of refinement is expected to hold. In addition, because the experiments used the ChatGPT web interface, the study did not provide direct control over temperature, random seed, or a fixed API-level model identifier, and the backend model may change over time. Therefore, the reported schedules should be interpreted as feasibility results obtained under the described interface conditions rather than as fully deterministic outputs expected to be exactly reproducible from the same prompts. We therefore do not claim broad prompt generalizability or deterministic schedule reproducibility from the present single-institution simulation. Future work should evaluate whether core prompt templates can be standardized, how much site-specific tuning is required, and whether prompt refinement can be made more reproducible through predefined development protocols, prospective failure-mode logging, repeated runs, complete input-output logging, and independent testing across multiple clinical scheduling environments. A more rigorous reproducibility study should use an API-based implementation with archived prompts and inputs, fixed model versions where available, controlled generation parameters, and systematic analysis of schedule stability across repeated executions.

\subsection{Limitations and Future Work}

This study has several limitations. {\revision First, the evaluations included a simulation study and a retrospective clinical comparison. These studies could not capture all clinical and operational constraints arising in a dynamic clinical environment. The findings therefore demonstrate the potential of the LLM-assisted scheduling framework, while prospective evaluation is needed to assess its practical workflow benefits.}

 Our immediate next step involves evaluating this framework as a human-in-the-loop scheduling interface through our secure, institutionally hosted LLM platform. This private environment ensures that scheduling data remain within the institutional firewall, addressing patient privacy, IT security, and data-governance barriers that typically preclude the use of public LLMs for clinical scheduling data. Because RT scheduling is commonly embedded within proprietary oncology information systems, the near-term goal is not autonomous writing of LLM-generated schedules into these systems. A more realistic first implementation would be a standalone decision-support tool that generates candidate schedules for therapist or scheduling-coordinator review. In this envisioned workflow, clinical staff would express scheduling priorities, constraints, and exception-handling requests in natural language; the LLM would generate candidate schedules or targeted revisions; and the resulting output would undergo deterministic validation, human review, and local approval before any accepted changes are implemented in the clinical scheduling system. A staged translation pathway would include retrospective testing with de-identified historical schedules, prospective shadow-mode evaluation without changing clinical schedules, usability and safety assessment with therapists and scheduling coordinators, institutional IT/security and data-governance review, and only later, if appropriate, controlled integration with oncology information systems through approved interfaces.

 Second, the comparison with optimization-based approaches should be interpreted as a preliminary benchmark rather than a comprehensive comparison. We implemented one optimization model and solved it using two algorithms, providing an initial reference for interpreting the LLM-generated schedules under the same simulated conditions. The optimization methods were not exhaustively tuned, and CP-SAT optimality status was not systematically analyzed. Future research should evaluate additional formulations, longer and matched solver budgets, documented optimality status, and broader objective-weighting strategies.

 Finally, although the framework demonstrated the ability to identify infeasible scheduling requests and generate interpretable corrective suggestions, the current study did not evaluate human factors, user trust, workflow integration, or safety governance considerations. Clinical implementation should therefore be framed as human-in-the-loop operational decision support rather than autonomous scheduling. In such a workflow, the LLM would serve as an interactive scheduling aid that helps translate clinical priorities and operational constraints into candidate schedules, while expert schedulers retain responsibility for reviewing feasibility, resolving contextual conflicts, incorporating patient-specific considerations, and approving the final schedule. Formal validation procedures, auditing mechanisms, and institutional oversight would be required before any clinical use. Future research should also investigate specialized interfaces that facilitate human-in-the-loop oversight, a critical component for clinical adoption of AI techniques\citep{jafar2026towards, daye2022implementation}. Although LLMs may reduce the manual burden of schedule generation, clinical scheduling will continue to require expert review because human schedulers account for contextual factors that may not be fully represented in structured prompts, including patient preferences, staffing constraints, machine maintenance, and last-minute clinical changes. LLM-generated self-evaluation may help summarize scheduling rationale and flag potentially difficult constraints, but it should be treated as an interpretability aid rather than a substitute for independent deterministic validation.

\section{Conclusion}

 This study evaluated the feasibility of using an existing LLM to generate candidate RT treatment schedules satisfying predefined clinical and operational constraints. To achieve this aim, we developed a structured prompting framework that supplied the LLM with scheduling rules, patient and schedule data, and scenario-specific operational objectives. In a realistic simulated three-LINAC environment, the framework produced schedules that satisfied feasibility rules across multiple scenarios, including a weekly time-consistency baseline, LINAC continuity, gap-constrained temporal relaxation, and infeasible scheduling request handling. We also compared LLM-generated schedules with optimization-based scheduling results to provide an initial benchmark for performance interpretation. It was observed that the LLM schedules achieved a clinically interpretable balance across mean gap, LINAC continuity, treatment-time consistency, and overtime, while the optimization-based schedules demonstrated expected trade-offs that depended on the solver and objective-weight profile. {\revision The six-week retrospective clinical comparison further demonstrated the potential to improve schedule compactness and treatment-time consistency at routine workload, with tradeoffs reflecting the selected scheduling objectives.} The findings suggested the potential value of LLM-assisted scheduling in providing a flexible natural-language, human-in-the-loop decision-support interface for translating clinical priorities into feasible candidate schedules. 

\section*{Acknowledgement}
During the preparation of this manuscript, the authors used ChatGPT by OpenAI to refine the language to enhance the general readability of the manuscript. The authors reviewed, edited, and verified the final text, and take full responsibility for the accuracy and integrity of the content.
\section*{References}
\addcontentsline{toc}{section}{\numberline{}References}

\end{document}